\documentclass[showpacs,twocolumn,prl,aps]{revtex4}
\usepackage{graphicx}

\begin{document}

\title[Short title for running header]{Spinon-Holon Recombination in Gutzwiller Projected Wave Functions}
\author{Hong-Yu Yang}
\affiliation{Center for Advanced Study, Tsinghua University,
Beijing, 100084, P. R. China}
\author{Tao Li}
\affiliation{Department of Physics, Renmin University of China,
Beijing, 100872, P. R. China}
\date{{\small \today}}

\begin{abstract}
The Gutzwiller projection is shown to induce nontrivial correlation
between the spinon and the holon in the slave Boson theory of the
$t-J$ model. We find this correlation is responsible some subtle
differences between the slave Boson mean field theory and the
Gutzwiller projected wave function(GWF), among which the
particle-hole asymmetry in the quasiparticle weight calculated from
the GWF is a particular example.
\end{abstract}

\maketitle

Superconductivity results from Bose condensation of charged
particles. In the BCS theory of superconductivity, Fermionic
electrons are paired into Bosonic Copper pairs which then condense
into a superfluid. Soon after the discovery of high temperature
superconductors, Anderson proposed an exotic way toward
superconductivity in this class of materials. Anderson's proposal
for superconductivity is to fractionalize the electron, rather than
pair them up\cite{Anderson,Baskran,Zou}. According to Anderson's
argument, doping holes into the parent compounds of the high
temperature superconductors, which are two dimensional
antiferromagnetic insulators, would generate a liquid-type spin
state called RVB state. Anderson argued that the excitations above
the RVB state are fractionalized. More specifically, the spin and
the charge quantum number of the electron are carried separately by
two kinds of excitations, namely a spin-$\frac{1}{2}$ Fermionic
excitation called spinon and a charge 1 Bosonic charged excitation
called holon. In such a spin-charge separated system, the charged
holon is liberated from the Fermionic statistics of the electron and
is ready to condense into a superfluid.

A problem with Anderson's proposal is that the predicted $T_{c}$ is
too high\cite{Kotliar,Lee1}. The holon supercurrent is not
efficiently dissipated in this mechanism. In the BCS theory of
superconductivity, the supercurrent is dissipated by thermally
generated quasiparticle excitations. These quasiparticles, which are
charged Fermionic excitations, form a normal fluid and cause
dissipative response in external electromagnetic(EM) field. On the
other hand, in an ideal spin-charge separated system, the Fermionic
spinon has been deprived of the charge to cause dissipative response
in an EM field, while the Bosonic excitation of the holon system is
very inefficient in dissipating the supercurrent.

One suggested solution for the above problem is to recombine the
separated spinon and holon\cite{Lee2,Wen,Ng}. By spinon - holon
recombination, the Fermionic spinon excitation acquire the charge to
cause dissipation in an EM field, or, the charged holon regain the
Fermionic statistics to be transformed into a normal carrier outside
the condensate.

However, the origin and even the exact meaning of such recombination
are still elusive. In this paper, we address the problem of
spinon-holon recombination in the slave Boson theory of the $t-J$
model. The slave Boson theory is the most direct way to realize the
RVB idea. In the slave Boson theory, the constrained electron
operator $\hat{c}_{i\sigma}$ is rewritten as
$f_{i\sigma}b_{i}^{\dagger}$, in which $f_{i\sigma}$ denotes the
Fermionic spinon and $b_{i}^{\dagger}$ denotes the Bosonic holon. To
be a faithful representation of the original electron, the slave
particles should be subjected to the local constraint
$\sum_{\sigma}f_{i,\sigma}^{\dagger}f_{i,\sigma}+b_{i}^{\dagger}b_{i}=1$.

At the mean field level, the local constraint on the slave particle
is relaxed to a global one. The spinon and the holon then move
independently. Wen and Lee proposed that the Gaussian fluctuation
around the mean field solution, which are gauge fluctuations, tends
to recombine the spinon and the holon\cite{Wen}. However, the lack
of a reliable treatment of gauge fluctuation in 2+1 dimension make
it hard to make any quantitative prediction.

In this paper, we address the problem of spinon-holon recombination
in the Gutzwiller projected wave functions(GWF). The GWF is derived
from the slave Boson mean field state by filtering out its
unphysical components with doubly occupied sites. Technically, the
Gutzwiller projection, which enforces the local constraint, accounts
partially the effect of gauge fluctuation. Thus some kind of
spinon-holon recombination is expected in GWF. We find the
Gutzwiller projection not only enforces the local constraint, but
also induces longer range correlation between the spinon and the
holon. This correlation is given by a well defined correlation
function and is found to be responsible for some subtle differences
between the mean field theory and the GWF. For example, the
particle-hole asymmetry in the quasiparticle weight calculated from
GWF is shown to be due to this correlation.

We start from the $t-J$ model which is defined as
\begin{equation}
    \mathrm{H}=-t\sum_{<i,j>,\sigma}(\hat{c}_{i,\sigma}^{\dagger}\hat{c}_{j,\sigma}+h.c.)
    +J\sum_{<i,j>}(\mathbf{S}_{i}\mathbf{S}_{j}-\frac{1}{4}n_{i}n_{j}),
\end{equation}
in which
$\mathbf{S}_{i}=\sum_{\alpha\beta}\hat{c}_{i,\alpha}^{\dagger}\mathbf{\sigma}_{\alpha\beta}\hat{c}_{i,\beta}$
and
$n_{i}=\sum_{\alpha}\hat{c}_{i,\alpha}^{\dagger}\hat{c}_{i,\alpha}$.
The electron operator $\hat{c}_{i,\sigma}$ in (1) is subjected to
the constraint of no boule occupancy
\begin{equation}
\sum_{\sigma}\hat{c}_{i,\sigma}^{\dagger}\hat{c}_{i,\sigma}\leq 1.
\end{equation}

In terms of the spinon and the holon operator, the $t-J$ model
reads\cite{Kotliar}
\begin{eqnarray*}
   \mathrm{H} &=& -t\sum_{<i,j>,\sigma}(f_{i,\sigma}^{\dagger}f_{j,\sigma}b_{j}^{\dagger}b_{i}+h.c.) \\
   &+&\frac{J}{2}\sum_{<i,j>,\alpha,\beta}(f_{i\alpha}^{\dagger}f_{i\beta}f_{j\beta}^{\dagger}f_{j\alpha}
   -f_{i\alpha}^{\dagger}f_{i\alpha}f_{j\beta}^{\dagger}f_{j\beta}),
\end{eqnarray*}
in which $f_{i\alpha}$ and $b_{i}$ are now subjected to the
constraint
\begin{equation}
\sum_{\sigma}f_{i,\sigma}^{\dagger}f_{i,\sigma}+b_{i}^{\dagger}b_{i}=1.
\end{equation}

In the mean field theory, the interaction term is decoupled by
introducing RVB order parameter $\chi_{ij}=\sum_{\sigma}\langle
f_{i,\sigma}^{\dagger}f_{j,\sigma} \rangle$ and
$\Delta_{ij}=\sum_{\alpha}\langle
\epsilon_{\alpha\beta}f_{i,\alpha}^{\dagger}f_{j,\beta} \rangle$, in
which $\epsilon_{\alpha\beta}$ is the total antisymmetric tensor.
The local constraint is relaxed to a global one. The mean field
Hamiltonian for the spinon and the holon take the form
\begin{equation}
    \mathrm{H}_{f}=\sum_{\mathrm{k},\sigma}\xi_{\mathrm{k}}f_{\mathrm{k},\sigma}^{\dagger}f_{\mathrm{k},\sigma}+
\sum_{\mathrm{k},\sigma}\Delta_{\mathrm{k}}
(f_{\mathrm{k},\sigma}^{\dagger}f_{\mathrm{-k},\bar{\sigma}}^{\dagger}+h.c.)
\end{equation}
and
\begin{equation}
    \mathrm{H}_{h} =\sum_{\mathrm{k}}\epsilon_{\mathrm{k}}b_{\mathrm{k}}^{\dagger}b_{\mathrm{k}}
\end{equation}
in momentum space. The mean field ground state is given by the
following product
\begin{equation}
  |\mathrm{G}\rangle_{\mathrm{MF}}
  =(b_{q=0}^{\dagger})^{N_{b}}|\mathrm{BCS}\rangle,
\end{equation}
in which $|\mathrm{BCS}\rangle$ denotes the BCS pairing state of the
spinon. A mean field excitation, for, say, a hole-like quasipaticle,
is given by
\begin{equation}
  |\mathrm{k},\sigma\rangle_{\mathrm{MF}}
  =b_{q=0}^{\dagger}\gamma_{\mathrm{k}\sigma}^{\dagger}|\mathrm{G}\rangle_{\mathrm{MF}},
\end{equation}
in which $\gamma_{\mathrm{k}\sigma}^{\dagger}$ denotes the Bogliubov
quasiparticle of the BCS Hamiltonian for spinon.

The GWF for the ground state is given by projecting
$|\mathrm{G}\rangle_{\mathrm{MF}}$ into the subspace that satisfy
the constraint Eq.(3),
\begin{equation}
  |\mathrm{G}\rangle_{\mathrm{Var}}
  =\mathrm{P_{G}}(b_{q=0}^{\dagger})^{N_{b}}|\mathrm{BCS}\rangle,
\end{equation}
in which $|\mathrm{G}\rangle_{\mathrm{Var}}$ denotes the variational
ground state and $\mathrm{P_{G}}$ denotes the Gutzwiller projection.
Similarly, $\mathrm{P_{G}}|\mathrm{k}\rangle_{\mathrm{MF}}$ provides
a variational guess for the quasipaticle excitation on
$|\mathrm{G}\rangle_{\mathrm{Var}}$. The same construction for the
quasiparticle excitation is used in some recent works\cite{Yunoki}.

The spinon-holon recombination in GWF can be most easily seen by
inspecting the quasiparticle weight $\mathrm{Z_{k}}$. In the mean
field theory, $\mathrm{Z_{k}}$ scales linearly with hole density ,
since the coherent spectral weight is caused by holon condensation.
However, $\mathrm{Z_{k}}$ calculated from GWF vanishes more slowly
near half filling. For example, $\mathrm{Z_{k}}$ calculated from the
one dimensional GWF vanishes as $x^{1/2}$ near half
filling\cite{Gebhard}. Similar behavior is also observed in two
dimensional systems\cite{Yang}. In some cases, $\mathrm{Z_{k}}$
calculated from GWF can even be nonzero at half filling\cite{TKLee}.
This discrepancy can be naturally explained by invoking spinon-holon
recombination in GWF.

Now we define spinon-holon recombination in GWF more explicitly. For
clarity's sake, we first consider a half filled system and assume
the spinon pairing term to be absent\cite{note1}. The mean field
ground state of the system is then given by a half filled spinon
Fermi sea, while a hole-like quasiparticle is generated as follow
\begin{equation}
    |\mathrm{k},\uparrow
    \rangle_{\mathrm{MF}}=b_{\mathrm{q=0}}^{\dagger}f_{-\mathrm{k},\downarrow}|\mathrm{FS}\rangle,
\end{equation}
in which $|\mathrm{FS}\rangle$ denotes the half filled Fermi sea. At
the mean field level, $|\mathrm{k},\uparrow\rangle_{\mathrm{MF}}$ is
interpreted as a state with an added holon and a hole of spinon on
the background of $|\mathrm{FS}\rangle$. In real space,
\begin{equation}
    |\mathrm{k},\uparrow\rangle_{\mathrm{MF}}=\frac{1}{N}\sum_{i,j}e^{i\mathrm{k}(r_{i}-r_{j})}b_{i}^{\dagger}f_{j\downarrow}|\mathrm{FS}\rangle.
\end{equation}
Thus, the added holon and the hole of spinon move independently on
$|\mathrm{FS}\rangle$ and have a chance of only $\frac{1}{N}$ to
recombine into the original electron. This explains the vanishing of
the quasiparticle weight at half filling in the mean field theory.

Now let's see what happens when the local constraint is enforced by
the Gutzwiller projection. The projected wave function for the above
quasiparticle is given by
\begin{equation}
    |\mathrm{k},\uparrow\rangle_{\mathrm{Var}}
    =\frac{1}{N}\sum_{i,j}e^{i\mathrm{k}(r_{i}-r_{j})}\mathrm{P_{G}} b_{i}^{\dagger}f_{j\downarrow}|\mathrm{FS}\rangle.
\end{equation}
Following the convention of the mean field theory, we interpret the
index $i$ and $j$ as the locations of the added holon and the hole
of spinon. The difference between
$|\mathrm{k},\uparrow\rangle_{\mathrm{Var}}$ and
$|\mathrm{k},\uparrow\rangle_{\mathrm{MF}}$ can be understood as
follows. In the mean field theory, the relative motion between the
added holon and the hole of spinon is described by plane wave
factor. The background particles in $|\mathrm{FS}\rangle$ do not
contribute to their correlation. After the Gutzwiller projection,
thees added particles get correlated with the background particles
through the local constraint. Thus the background particles in
$|\mathrm{FS}\rangle$ can also contribute to the correlation between
the added holon and the hole of spinon in
$|\mathrm{k},\uparrow\rangle_{\mathrm{Var}}$.

First, we assume $i \neq j$. Due to the local constraint, spinons in
$|\mathrm{FS}\rangle$ must keep away from the site $i$, which is
already assigned to the added holon. At the same time, site $j$
should be doubly occupied in $|\mathrm{FS}\rangle$, since it
accommodates a spin in the projected state. The sites other than $i$
and $j$ should be all singly occupied in $|\mathrm{FS}\rangle$.
Thus, the probability of finding the holon at site $i$ while the
hole of spinon at site $j$, which is denoted as $P_{ij}$, is given
by that of finding site $i$ empty, site $j$ doubly occupied, and all
other sites singly occupied in $|\mathrm{FS}\rangle$. For $i=j$, in
which case a real electron is removed from site $i$, the
corresponding probability($P_{ii}$) is given by that of finding site
$i$ occupied by a down spin and all other sites singly occupied in
$|\mathrm{FS}\rangle$. The ratio between $P_{ij}$ and $P_{ii}$ is
given by
\begin{equation}
\frac{P_{ij}}{P_{ii}}=\frac{\sum_{\beta}|\psi_{\beta}|^{2}}{\sum_{\alpha}|\psi_{\alpha}|^{2}}=
   \frac{\sum_{\alpha}|\psi_{\alpha}|^{2}
   \frac{|\psi_{\beta}|^{2}}{|\psi_{\alpha}|^{2}}}{\sum_{\alpha}|\psi_{\alpha}|^{2}}.
\end{equation}
Here, $\psi_{\alpha}$ denotes the amplitude of $|\mathrm{FS}\rangle$
for a general configuration $|\alpha\rangle$ in the subspace of no
double occupancy, $\psi_{\beta}$ denotes the amplitude for the
configuration derived from $|\alpha\rangle$ by moving an electron(of
either spin) from site $i$ to site $j$. This statistical sum can be
evaluated easily with Variational Monte Carlo method.

To see the value of the thus defined correlation function, we
consider some special cases in which the effect of spinon-holon
recombination is manifest. As a trivial example, we consider the
case of removing an electron from a fully polarized spin background.
According to the slave Boson mean field theory, the quasipaticle
weight is given by the hole density and vanishes at half filling.
However, since the spin is fully polarized, the state in
consideration is in fact a mean field state in which the
quasiparticle weight should be exactly one. This discrepancy can be
easily resolved by noting that $P_{ij}$ is nonzero only for $i=j$ on
a fully polarized spin background. This indicates that the added
holon and the hole of spinon must occupy the same site and thus
recombine into a real electron on that site.

As a less trivial example, we consider a state with
antiferromagnetic long range order. As we will show below, $P_{ij}$
decays exponentially in this case. The exponential decay of $P_{ij}$
can be understood by noting the opening of a SDW gap between
configuration $|\beta\rangle$ and $|\alpha\rangle$ in Eq. (12). Thus
the added holon and the hole of spinon will form a well defined
bound state in such a spin background. Since the bound state has a
nonzero overlap with a bare electron, the quasiparticle weight
should be finite even at half filling. Calculation in \cite{TKLee}
does find a finite quasiparticle weight for such a state.

\begin{figure}[h!]
\includegraphics[width=8cm,angle=0]{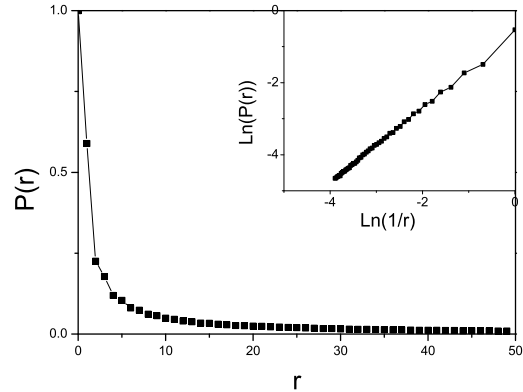}
\caption{Spinon - holon correlation function for the one dimensional
projected Fermi sea. The inset show the data in logarithmic scale.}
\label{fig1}
\end{figure}

Now we present the numerical results. Figure 1 shows the correlation
function for the one dimensional projected Fermi sea. The
correlation function decays as $1/r$ at large distance and is
unnormalizable. Thus the quasiparticle weight should vanish at half
filling. The result for the two dimensional d-wave RVB state is
shown in Figure 2. The correlation function now decays as
$1/r^{\frac{3}{2}}$ at large distance, which is also unnormalizable.
Thus the quasiparticle weight should also vanish at half filling in
this state. In Figure 3, we show the result for the antiferromagetic
ordered state. As mentioned above, the correlation function decays
exponentially in this case. However, the short range behavior is
quite similar to that of the d-wave RVB state.

\begin{figure}[h!]
\includegraphics[width=9cm,angle=0]{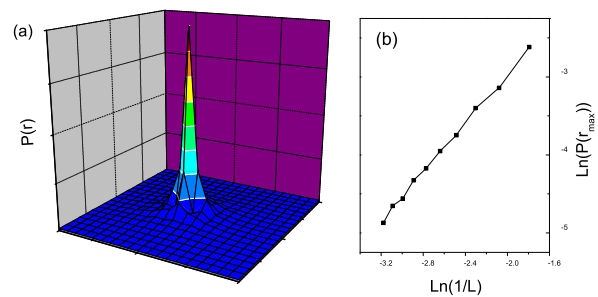}
\caption{(a)Spinon - holon correlation function for a d-wave RVB
state. The d-wave RVB order parameter $\Delta/t=0.25$. The
calculation is done on a $20\times20$ lattice. (b)Power law decay of
the spinon - holon correlation function. L denotes the linear size
of the lattice, $r_{\mathrm{max}}$ is largest distance that can be
defined on a periodic lattice of size L.} \label{fig2}
\end{figure}

\begin{figure}[h!]
\includegraphics[width=9cm,angle=0]{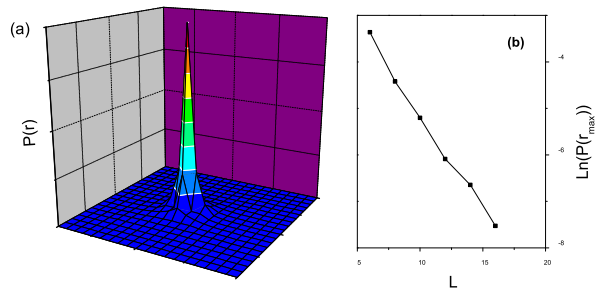}
\caption{(a)Same as Figure 2, but now for an antiferromagnetic
ordered state. The SDW and d-wave order parameters are
$\Delta_{AF}/t=0.1$ and $\Delta/t=0.25$. (b)Exponential decay of the
spinon - holon correlation function.} \label{fig3}
\end{figure}

The spinon-holon recombination discussed above refers to the
correlation between the added particles rather than that between the
background particles. At finite doping, a holon condensate is
established and the background spinon has a chance of $x$ to behave
as a coherent quasipaticle. In fact, the latter is the only way that
the system can build a finite quasiparticle weight on the particle
side of the spectrum. Following essentially the same steps detailed
above, one easily sees that the added slave paricles(or hole of
them) during the process of injecting an electron have no
correlation with each other. Since the quasiparticle weight due to
holon condensation is by definition particle-hole symmetric, the
particle-hole asymmetry in the quasiparticle weight should be
attributed to the spinon-holon recombination effect discussed in
this paper\cite{Yang}.

Especially, while the quasiparticle weight for adding an electron is
constrained by the sum rule to vanish at half filling, the
quasiparticle weight for removing an electron can be finite at half
filling, provided that the added holon and the hole of spion form
well defined bound state, as is the case in the antiferromagnetic
ordered state. In such a case the quasiparticle can contribute a
substantial part of the tunneling asymmetry, although for a full
understanding of the latter one also should take into account the
contribution from incoherent spectral weight\cite{Anderson1}. When
the added holon and the hole of spinon are less tightly bounded, the
particle-hole asymmetry in the quasiparticle weight should be less
dramatic, as is found in a recent work on the tunneling asymmetry of
high temperature superconductors\cite{Yang}.

We now discuss some other effects of spinon-holon recombination. By
recombining with a spinon excitation, a hole changes the local spin
environment around it. This effect can be invoked to release the
kinetic energy of the holes in certain spin background through
spontaneous generation of spinon excitations in the system. An
example in this respect is provided by a recent variational study on
the slightly doped cuprates\cite{TKLee}. In that work, a $t-J$ model
with next-nearest-neighboring hoping term is considered. At small
doping , it is found that a variational state with spontaneously
generated spinon excitations give lower energy than a state without
spinon excitations. It is also found that the presence of these
spinon excitations changes the charge correlation dramatically. This
can also be understood in terms of spinon-holon recombination
defined in this paper. A detailed analysis of these issues will
appear in separate paper.

Finally, we revisit the problem of supercurrent dissipation. In the
presence of spinon excitations, a holon tend to bind with them and
form a Fermionic composite object. A holon is thus transformed into
a normal charge carrier out of its condensate. The superconductivity
is gone when all holons in the condensate are transformed into
normal carriers. At low doping, a small number of spinon excitation
is enough to kill the superconductivity which is not expected to
alter the spin correlation significantly. This is proposed as a
mechanism for the pseudogap in underdoped cuprates\cite{Lee2,Wen}.

In conclusion, an explicit definition for spinon-holon recombination
is given on the GWF. The thus defined spinon-holon recombination is
shown to be responsible for the particle-hole asymmetry in the
quasiparticle weight calculated from the GWF and some other subtle
differences between the mean field theory and the GWF. A new
mechanism for the dissipation of supercurrent in the RVB state is
proposed based on the spinon-holon recombination defined in this
paper.

This work is supported by NSFC Grant No.90303009.


\begin{thebibliography}{99}
\bibitem {note1}One easily verifies that the only effect of a nonzero pairing
term is to replace $|\mathrm{FS}\rangle$ with $|\mathrm{BCS}\rangle$
in the following discussion, provided that a cannonical ensamble is
used.
\end{thebibliography}
\end{document}